\documentclass[prd,showpacs]{revtex4}

\usepackage{amsmath}
\usepackage{amsthm}
\usepackage{amssymb}
\usepackage{amscd}
\usepackage[british]{babel}
\usepackage{graphicx}
\usepackage{psfrag}
\usepackage{epsfig}
\usepackage{rotating}
\usepackage{times}

\begin{document}

\title{Black hole collapse simulated by vacuum fluctuations with moving semi-transparent mirror}

%The spectrum of radiation emitted by a partially transmitting mirror following
%prescribed trajectories simulating a black hole collapse}

\author{Jaume Haro$^{1,}$\footnote{E-mail: jaime.haro@upc.edu} and
Emilio Elizalde$^{2,}$\footnote{E-mail: elizalde@ieec.uab.es,
elizalde@math.mit.edu}}

\affiliation{$^1$Departament de Matem\`atica Aplicada I, Universitat
Polit\`ecnica de Catalunya, Diagonal 647, 08028 Barcelona, Spain \\
$^2$Instituto de Ciencias del Espacio (CSIC) \& Institut
d'Estudis Espacials de Catalunya (IEEC/CSIC)\\ Campus UAB, Facultat
de Ci\`encies, Torre C5-Parell-2a planta, 08193 Bellaterra
(Barcelona) Spain}

%\author{Jaume Haro and Emilio Elizalde}

\date{November 6, 2007}

\pagestyle{myheadings}

\theoremstyle{plain}
\newtheorem{lemma}{Lemma}[section]
\newtheorem{theorem}{Theorem}[section]
\newtheorem{proposition}{Proposition}[section]
\newtheorem{corollary}{Corollary}[section]
\newtheorem{remark}{Remark}[section]
\newtheorem{definition}{Definition}[section]
\newtheorem{example}{Example}[section]

\newcommand{\boxend}{\flushright{$\Box$}}
\newenvironment{dem}[1]{\begin{trivlist} \item {\bf Proof #1\/:}}
            {\boxend \end{trivlist}}

\newcommand{\N}{{\mathbb N}}               % for natural numbers
\newcommand{\Z}{{\mathbb Z}}               % for entire numbers
\newcommand{\Q}{{\mathbb Q}}               % for rational numbers
\newcommand{\R}{{\mathbb R}}               % for real numbers
\newcommand{\C}{{\mathbb C}}               % for complex numbers
\renewcommand{\S}{{\mathbb S}}             % for the unity circle
\newcommand{\T}{{\mathbb T}}               % for the standard torus
\newcommand{\D}{{\mathbb D}}               % for the dihedral groups

\newcommand{\half}{\frac{1}{2}}
\renewcommand{\Re}{\mbox{\rm Re}}
\renewcommand{\Im}{\mbox{\rm Im}}
\newcommand{\sprod}[2]{\left\langle#1,#2\right\rangle}
\newcommand{\inner}[2]{\left\langle#1,#2\right\rangle_2}

\newcommand{\e}{\epsilon}
\newcommand{\w}{\omega}
\newcommand{\f}{\frac}

\newcommand{\ep}{\varepsilon}
\newcommand{\al}{\alpha}
\newcommand{\h}{\hbar}
\renewcommand{\tilde}{\widetilde}

%\begin{document}

%\noindent
%Departament de Matem\`atica Aplicada I,
%Universitat Polit\`ecnica de Catalunya,
%\newline
%Diagonal 647, 08028 Barcelona, Spain.
%\newline
%e-mail:  jaime.haro@upc.es

%\noindent$^1$Departament de Matem\`atica Aplicada I, Universitat
%Polit\`ecnica de Catalunya, Diagonal 647, 08028 Barcelona, Spain \\
%$^2$Instituto de Ciencias del Espacio (CSIC) \& Institut d'Estudis
%Espacials de Catalunya (IEEC/CSIC)\\ Campus UAB, Facultat de
%Ci\`encies, Torre C5-Parell-2a planta, 08193 Bellaterra (Barcelona)
%Spain

\thispagestyle{empty}

\begin{abstract}

Creation of scalar massless particles in  two-dimensional Minkowski
space-time---as predicted by the dynamical Casimir effect---is
studied for the case of a semitransparent mirror initially at rest,
then accelerating for some finite time, along a trajectory that
simulates a black hole collapse (defined by Walker, and Carlitz
and Willey), and finally moving with constant velocity. When the
reflection and transmission coefficients are those in the model
proposed by Barton, Calogeracos, and Nicolaevici
[$r(w)=-i\alpha/(\w+i\alpha)$ and $s(w)=\w/(\w+i\alpha)$, with
$\alpha\geq 0$], the Bogoliubov coefficients on the back side of the
mirror can be computed exactly. This allows us to prove that,  when
$\alpha$ is very large (case of an ideal, perfectly reflecting
mirror) a thermal emission of scalar massless  particles obeying
Bose-Einstein statistics is radiated from the mirror (a black body
radiation), in accordance with results previously obtained in the literature.
However, when $\alpha$ is finite (semitransparent mirror, a
physically realistic situation) the striking result is obtained that
the thermal emission of scalar massless particles obeys Fermi-Dirac
statistics. We also show here that the reverse change of statistics
takes place in a bidimensional fermionic model for massless
particles, namely that the Fermi-Dirac statistics for the completely
reflecting situation will turn into the Bose-Einstein statistics for
a partially reflecting, physical mirror.

\end{abstract}


\maketitle

\vspace{1cm}

%{\bf Keywords.} Partially Transmitting Moving Mirrors, Radiation
%Spectrum, Particle Creation, Black hole Collapse.

\vspace{0.5cm}

%{\bf PACS 2003 subject classification.} 03.70.+k, 04.62.+v

%\clearpage

%\tableofcontents

%\clearpage

%\markboth{The spectrum of radiation emitted by ...} {J. Haro and E.
%Elizalde}

\section{Introduction}

The Davies-Fulling model \cite{fd76,fd77} describes the creation of
scalar massless particles by a moving perfect mirror following a
prescribed trajectory. This phenomenon is also termed as the
dynamical Casimir effect. Recently, the authors of the present paper
introduced a Hamiltonian formulation in order to address some
problems associated with the physical description of this effect in
the time interval while the mirror is moving \cite{he06,he06a}; in
particular, of the regularization procedure, which turns out to be
decisive for the correct derivation of physically meaningful
quantities. A basic difference with previous results was that the
motion force derived within the new approach contains a reactive
term---proportional to the mirror's acceleration. This term is of
the essence in order to obtain particles with a positive energy all
the time while the oscillation of the mirror takes place, and which
always satisfy the energy conservation law. Those result followed
essentially from the introduction of physically realistic
conditions, e.g. a semi-transparent or partially transmitting
mirror, which is perfectly reflecting for low frequencies but
becomes transparent to very high ones.

Here we will study a different aspect of the introduction of
physically plausible, semitransparent mirrors, namely the particle
spectrum produced---in the conditions of the Fulling-Davies
effect---by a mirror of this sort which is initially at rest, then
accelerates during a large enough (but finite) time span, $u_0$,
along a trajectory that simulates a black hole collapse, as defined
by Walker \cite{w85}, and Carlitz and Willey \cite{cw87}:
\begin{eqnarray} v=\frac{1}{k}(1-e^{-ku})\end{eqnarray} (in light-like
coordinates, where $k$ is some frequency), and finally, for $u\geq
u_0$, is left alone moving with constant velocity in an inertial
trajectory.

We will be interested in calculating the radiation emitted by the
mirror from its back (e.g. right) side. As is well-known,  a perfect
mirror that follows this kind of trajectory produces a thermal
emission of scalar massless particles obeying Bose-Einstein
statistics. More precisely, for $1\ll\w'/k\ll e^{ku_0}$  and
$1\ll\w'/\w\ll e^{ku_0}$, one has \cite{c02b,n03,ha05}
\begin{eqnarray}
\left|\beta_{\omega,\omega'}^{R,R}\right|^2\equiv\left|
({\phi_{\omega,R}^{out}}^*;\phi_{\omega',R}^{in})\right|^2
\cong\frac{1}{2\pi\omega' k}
\left(e^{2\pi\omega/ k}-1\right)^{-1}.
\end{eqnarray}
Turning to the case of a partially reflecting mirror---in which we
will be mainly interested in this paper---in order to obtain the
radiation on its right hand side (rhs), we also need
to calculate the Bogoliubov coefficient:
$\beta_{\omega,\omega'}^{R,L}\equiv{({\phi_{\omega,R}^{out}}^*; \phi_{\omega',L}^{in})}^*$.

We thus first obtain the `in' modes on the rhs of the mirror when
the reflection and transmission coefficients are
$r(w)=\frac{-i\alpha}{\w+i\alpha}$ and $s(w)=\frac{\w}{\w+i\alpha}$,
with $\alpha\geq 0$, that is, when the Lagrangian density is given
by \cite{bc95,n01,c02a}
\begin{eqnarray}
{\mathcal L}=
\frac{1}{2}[(\partial_t\phi)^2-(\partial_z\phi)^2]-\alpha
\sqrt{1-\dot{g}^2(t)}\phi^2\delta(z-g(t)),
\end{eqnarray}
where $z=g(t)$ is the trajectory in the $(t,z)$ coordinates.

\subsection{The main results} We now state the main results that will be obtained
in this paper (for $1\ll\w'/k\ll e^{ku_0}$ and $1\ll\w'/\w\ll
e^{ku_0}$).
\begin{enumerate}\item For a perfectly reflecting mirror, i.e.,
when $\w'\ll\alpha$, we will see that
\begin{eqnarray}\left|\beta_{\omega,\omega'}^{R,R}\right|^2 \cong\frac{1}{2\pi\omega' k}
\left(e^{2\pi\omega/k}-1\right)^{-1}, \quad
\left|\beta_{\omega,\omega'}^{R,L}\right|^2\cong 0,
\end{eqnarray}
namely, a thermal radiation of massless particles obeying
 Bose-Einstein statistics is produced.
\item For a perfectly transparent mirror, i.e.,when $\alpha\cong 0$, we will get that
\begin{eqnarray} |\beta_{\omega,\omega'}^{R,R}|^2 \cong 0,
\quad |\beta_{\omega,\omega'}^{R,L}|^2\cong 0.
\end{eqnarray}
In other words, there is no  particle production.
\item  In the physically more realistic case of a partially transmitting mirror
(transparent to high enough frequencies \cite{he06}), i.e., when
$\alpha\ll \w'$, what we obtain is
\begin{eqnarray}\label{A} && \left|\beta_{\omega,\omega'}^{R,R}\right|^2
\cong\frac{1}{2\pi\omega k}\left(\frac{\alpha}{\w'}\right)^2
\left(e^{2\pi\omega/ k}+1\right)^{-1}, \nonumber \\
&& \left|\beta_{\omega,\omega'}^{R,L}\right|^2\sim \frac{1}{\w\w'}
{\mathcal O}\left[ \left(\frac{\alpha}{\w'}\right)^2\right].
\end{eqnarray}
And, since $\left|\beta_{\omega,\omega'}^{R,L}\right|\ll
\left|\beta_{\omega,\omega'}^{R,R}\right|$, we then conclude that a
semitransparent mirror emits a thermal radiation of scalar massless
particles obeying Fermi-Dirac statistics.

Moreover, Eq.~(\ref{A}) does show that  there are no ultraviolet
divergences when the mirror is semi-transparent (in agreement with
previous conclusions in \cite{he06,he06a,12p}) and,
consequently, the number of produced particles in the $\w$ mode,
namely ${\mathcal N}_{\w}$, is finite. In fact, with good
approximation we have calculated that
\begin{eqnarray}
{\mathcal N}_\w\cong\frac{1}{2\pi \w}\left(\frac{\alpha}{k}\right)^2
\left(e^{2\pi\w/k}+1\right)^{-1}.
\end{eqnarray}
It must be here remarked that this phenomenon will in no way
occur for a perfectly reflecting mirror, where the number of produced
particles in a prescribed mode diverges linearly with time
\cite{n03,ha05}.
\end{enumerate}

Actually, the same kind of effect---but now reversed---occurs if a
bidimensional {\it fermionic} model of massless particles is
considered, where in the perfectly reflecting case the mirror emits
a thermal radiation of fermions obeying the Fermi-Dirac statistics.
When the mirror becomes semi-transparent (the physically realistic
case), the emitted thermal radiation will obey Bose-Einstein
statistics.
We interpret these results as a proof of the fact that the spectrum
of the radiation produced by a mirror which follows a trajectory
that simulates black hole collapse does {\it not} depend on the
statistics of the field, being just determined by the  interaction
of the mirror with the radiation field. In our case this interaction
is given by the reflection and transmission coefficients which
depend on the parameter $\alpha$, that determines the spectrum of
the emitted radiation.

Here it is important to emphasize that the word `statistics' refers
all the time to the $\beta$-Bogoliubov coefficient characterizing
the spectrum of the radiated particles and {\it not} to the algebra
obeyed by the creation and annihilation operators, that always satisfy
the corresponding canonical commutation relations (anti-commutation,
in the fermionic case). That is, e.g. in the second case studied
the original particles continue to be fermions, but the spectrum
of the radiated emission corresponds to bosonic ones, when the
mirror becomes physical, that is, semi-transparent. As a consequence,
fermionic number is not actually violated.

Related phenomenons of similar kind have been reported to occur
also in other situations:
\begin{enumerate}\item In the case of an electric
charge following the trajectory $v=\frac{1}{k}(1-e^{-ku})$. When the
radiation field has spin $1$, the radiation emitted by the charge
obeys Bose-Einstein statistics, but when a scalar charge, and
consequently an scalar radiation field, is considered, the emitted
radiation will obey  Fermi-Dirac statistics (see \cite{nr95} for
more details).
\item When measuring the spectrum of a scalar field by using a
DeWitt detector which follows a uniformly accelerated world-line
in Minkowski space-time, one can show that, when the dimension of
the space-time is even the Bose-Einstein statistics is obtained; however when
this dimension is odd the reverse change of statistics occurs
(see \cite{t86} for further details).
\end{enumerate}

Finally, in an Appendix we will specify some sufficient conditions in
order to ensure the convergence of the total number of produced
particles and of their associated energy.

\section{Perfectly reflecting, moving mirror}
Consider a massless scalar field $\phi$  in two-dimensional
Minkowski space-time interacting with a moving mirror. Assume that
the mirror trajectory simulates a black hole collapse
\cite{fd77,bd82}, this is, that it reduces to the following form in the
light-like coordinates $u\equiv t-z$ and $v\equiv t+z$:
\begin{eqnarray}
v=V(u)\equiv\left\{\begin{array}{ccc}
u&\mbox{if}& u\leq 0\\
&&\\
\frac{1}{k}(1-e^{-ku})&\mbox{if}& 0\leq u\leq u_0\\
&&\\
V(u_0)+A(u-u_0)&\mbox{if}& u\geq u_0,
\end{array}\right.\end{eqnarray}
with $A=e^{-ku_0}$, where $k$ is a frequency and $u_0\gg 1$.
Note that this trajectory can also be written as follows
\begin{eqnarray}
u=U(v)\equiv\left\{\begin{array}{ccc}
v&\mbox{if}& v\leq 0\\
&&\\
-\frac{1}{k}\ln(1-kv)&\mbox{if}& 0\leq v\leq v_0\\
&&\\
U(v_0)+A^{-1}(v-v_0)&\mbox{if}& v\geq v_0.
\end{array}\right.\end{eqnarray}
For a perfectly reflecting mirror, the sets of `in' and `out' mode
functions are \cite{dw75}
\begin{eqnarray}\left\{\begin{array}{c}
\phi_{\omega,R}^{in}(u,v)=\frac{1}{\sqrt{4\pi|\omega| }}
\left(e^{-i\omega v}-e^{-i\omega V(u)}\right)\theta(v-V(u))\\
\\
\phi_{\omega,L}^{in}(u,v)=\frac{1}{\sqrt{4\pi|\omega| }}
\left(e^{-i\omega u}-e^{-i\omega U(v)}\right)\theta(u-U(v))
\end{array}\right.
\end{eqnarray}
and
\begin{eqnarray}\left\{\begin{array}{c}
\phi_{\omega,R}^{out}(u,v)=\frac{1}{\sqrt{4\pi|\omega| }}
\left(e^{-i\omega u}-e^{-i\omega U(v)}\right)\theta(v-V(u))\\
\\
\phi_{\omega,L}^{out}(u,v)=\frac{1}{\sqrt{4\pi|\omega| }}
\left(e^{-i\omega v}-e^{-i\omega V(u)}\right)\theta(u-U(v)),
\end{array}\right.
\end{eqnarray}
respectively. Our main objective in this section will be to
calculate the beta Bogoliubov coefficient
\begin{eqnarray}
\beta^{R,R}_{\w,\w'}\equiv
{({\phi_{\omega,R}^{out}}^*;\phi_{\omega',R}^{in})}^*, \quad \mbox{
with } \quad \w,\w'>0,
\end{eqnarray}
where the brackets on the rhs denote the usual product for scalar
fields (see, e.g., Birrell-Davies \cite{bd82}). In order to compute
this coefficient we choose the right null future infinity domain,
${\mathcal J}_R^+$. There, we have
\begin{eqnarray}
\beta^{R,R}_{\w,\w'}=2i\int_{\R}du\phi_{\w,R}^{out}\partial_u\phi_{\w',R}^{in}&=&
\frac{1}{2\pi i\sqrt{\w\w'}}\frac{\w'}{\w+\w'}
     -\frac{1}{2\pi
i\sqrt{\w\w'}}e^{-i\w
u_0}e^{-i\w'V(u_0)}\frac{\w'A}{\w+\w'A}\nonumber\\&& -\frac{1}{2\pi k}
\sqrt{\frac{\w'}{\w}}
\int_{0}^{1-A}ds(1-s)^{i{\w}/{k}}e^{-i{\w'}/{k}s}.
\end{eqnarray}
Taking into account that  $1\ll {\w'}/{k}\ll A^{-1}$ and
$1\ll{\w'}/{k}\ll
 A^{-1}$, we get with good approximation
\begin{eqnarray}\label{eq1}
\beta^{R,R}_{\w,\w'}&\cong & \frac{1}{2\pi i\sqrt{\w\w'}}
      -\frac{1}{2\pi k} \sqrt{\frac{\w'}{\w}}
\int_{0}^{1-A}ds(1-s)^{i{\w}/{k}}e^{-i{\w'}/{k}s}.
\end{eqnarray}
To obtain an explicit expression for the second term on the rhs, we
consider the domain
\begin{eqnarray} D\equiv \{z\in \C |\Re\, z\in [0,1-A], \Im\, z \in
[-\epsilon,0], \mbox{ with } k/w'\ll \epsilon\ll 1 \}
\end{eqnarray} and going through the same steps as in \cite{ha05}, we
easily obtain
\begin{eqnarray}
\beta^{R,R}_{\w,\w'}\cong \frac{1}{2\pi i
\sqrt{\w\w'}}e^{-i{\w'}/{k}}\left(\frac{ik}{\w'}\right)^{i{\w}/{k}}
\Gamma\left(1+i{\w}/{k}\right).
\end{eqnarray}
As a consequence, using that
$|\Gamma\left(1+i{\w}/{k}\right)|^2=\frac{\pi{\w}/{k}}{\sinh\left(\pi{\w}/{k}
 \right)}$ (see \cite{as72}) we get the announced result, that for a
 perfect reflecting mirror the relevant Bogoliubov coefficient is
\begin{eqnarray}
\left|\beta^{R,R}_{\w,\w'}\right|^2\cong \frac{1}{2\pi \w'
k}\left(e^{2\pi{\w}/{k}}-1\right)^{-1}.
\end{eqnarray}

\section{Partially reflecting, moving mirror}

Now we start by reducing the problem to co-moving coordinates
$(\tau, \rho)$, that is, those for which the mirror remains at rest,
$\tau$ being the proper time of the mirror, and we take $\rho$ such
that its trajectory be given by $\rho=0$. Introducing the light-like
coordinates  $(\bar{u},\bar{v})$, defined as
\begin{eqnarray}
\bar{u}\equiv \tau-\rho;\quad \bar{v}\equiv \tau+\rho,
\end{eqnarray}
we will calculate the mirror's trajectory in the coordinates
$(\bar{u},\bar{v})$.  Along this trajectory, the length element
obeys the identity \cite{op01}
\begin{eqnarray}
d\tau^2=d\bar{u}^2=d\bar{v}^2=V'(u)du^2=U'(v)dv^2.
\end{eqnarray}
An easy calculation yields then the relations
\begin{eqnarray}
\bar{v}=\bar{u}(u)\equiv\left\{\begin{array}{ccc} u&\mbox{if}& u\leq 0\\
&&\\
\frac{2}{k}(1-e^{-k{u}/{2}})&\mbox{if}& 0\leq u\leq u_0\\
&&\\
\bar{u}(u_0)+\sqrt{A}(u-u_0)&\mbox{if}& u\geq u_0,
\end{array}\right.\end{eqnarray}
and
\begin{eqnarray}
\bar{u}=\bar{v}(v)\equiv\left\{\begin{array}{ccc}v&\mbox{if}& v\leq 0\\
&&\\
\frac{2}{k}(1-\sqrt{1-kv})&\mbox{if}& 0\leq v\leq v_0\\
&&\\
\bar{v}(v_0)+A^{-{1}/{2}}(v-v_0)&\mbox{if}& v\geq v_0.
\end{array}\right.\end{eqnarray}
When the semitransparent mirror is at rest,  scattering is described by the analytical
$S$-matrix (see \cite{he06a,jr91} for full details)
\begin{eqnarray}S(\w)=\left(\begin{array}{cc}
{s}(\w)&{r}(\w)e^{-2i\w L}\\
{r}(\w)e^{2i\w L}&{s}(\w)\end{array}\right),\end{eqnarray} where
$x=L$ is the position of the mirror. The $S$ matrix is taken to be
real in the temporal domain, causal, unitary, and the identity at
high frequencies \cite{jr91}, being $r(\omega)$ and $s(\omega)$ the
reflection and transmission functions, which are analytic and such
that tend in modulus to $-1$ and $0$, respectively, as $\omega \to 0$
(to $0$ and $1$, as $\omega \to \infty$). The `in' modes in the
coordinates $(\bar{u},\bar{v})$ are \cite{bc95}
\begin{eqnarray}\label{d}
g^{in}_{\w, R}(\bar{u},\bar{v})= \frac{1}{\sqrt{4\pi|\omega|
}}s(\w)e^{-i\omega
\bar{v}}\theta(\bar{u}-\bar{v})+\frac{1}{\sqrt{4\pi|\omega| }}
\left[e^{-i\omega \bar{v}}+r(\w)e^{-i\omega
\bar{u}}\right]\theta(\bar{v}-\bar{u}),
\end{eqnarray}
\begin{eqnarray}\label{e}
g^{in}_{\w, L}(\bar{u},\bar{v})=\frac{1}{\sqrt{4\pi|\omega| }}
\left[e^{-i\omega \bar{u}}+r(\w)e^{-i\omega
\bar{v}}\right]\theta(\bar{u}-\bar{v})+ \frac{1}{\sqrt{4\pi|\omega|
}}s(\w)e^{-i\omega \bar{u}}\theta(\bar{v}-\bar{u}).
\end{eqnarray}
Note that the `in' modes in the coordinates $(u,v)$, namely $\phi^{in}$, are defined
in the right null past infinity ${\mathcal J}^-_R$ by
\begin{eqnarray}
\phi^{in}_{\w,R}=\frac{1}{\sqrt{4\pi|\omega| }}e^{-i\w v}, \quad
\phi^{in}_{\w,L}=0,
\end{eqnarray}
and, in the left null past infinity ${\mathcal J}^-_L$, by
\begin{eqnarray}
\phi^{in}_{\w,R}=0, \quad
\phi^{in}_{\w,L}=\frac{1}{\sqrt{4\pi|\omega| }}e^{-i\w u}.
\end{eqnarray}

From these definitions it is clear that $\bar{g}^{in}_{\w,
k}(u,v)\equiv g^{in}_{\w, k}(\bar{u}(u),\bar{v}(v))$ with $k=R,L$
are not such modes. However, the modes $\bar{g}^{in}_{\w, k}$ do
actually constitute an orthonormal basis of the space of solutions
of our problem. As a consequence, if we use the fact that
$\bar{g}^{in}_{-\w, k}=\bar{g}^{in
*}_{\w, k}$, we can obtain the following relation
\begin{eqnarray}\label{a}
\phi^{in}_{\w,k}=\int_{\R}d\w'\chi(\w')(\bar{g}^{in}_{\w',
k};\phi^{in}_{\w,k})\bar{g}^{in}_{\w', k},\quad
k=R,L
\end{eqnarray}
with $\chi(\w')$  the sign function. To be remarked is that
Eq.~(\ref{a}) is to be interpreted as follows
\begin{eqnarray}
\phi^{in}_{\w,k}=\lim_{\lambda\rightarrow\infty}\int_{\R}d\w'\chi(\w')(\bar{g}^{in}_{\w',
k};\phi^{in}_{\w,k})\bar{g}^{in}_{\w', k}
F_{\lambda}(\w'),\end{eqnarray} being $F_{\lambda}(\w')$  a
frequency cut-off, as for instance $\frac{\lambda^2}{\lambda^2+(\w')^2}$.

To calculate the `in' modes explicitly, we have chosen the
coefficients
 $r(w)=\frac{-i\alpha}{\w+i\alpha}$ and $s(w)=\frac{\w}{\w+i\alpha}$
with $\alpha\geq 0$. In this case, on the rhs of the mirror we
obtain
\begin{eqnarray}
\phi^{in}_{\w,R}(u,v)=\frac{1}{\sqrt{4\pi|\omega| }}e^{-i\omega v}+
\phi_{\w, R}^{refl}(u);\qquad \phi^{in}_{\w,L}(u,v)=\phi_{\w,
L}^{trans}(u),
\end{eqnarray}
where
\begin{eqnarray*}\hspace{-2mm}
\phi_{\w, R}^{refl}(u)=\left\{
\begin{array}{cc}
\frac{1}{\sqrt{4\pi|\omega| }}\frac{-i\alpha}{\w+i\alpha}e^{-i\w V(u)};&u\leq 0\\
&\\
 \frac{1}{\sqrt{4\pi|\omega| }}\frac{-i\alpha}{\w+i\alpha}e^{-\alpha\bar{u}(u)}
 -\frac{2\alpha}{k\sqrt{4\pi|\omega|
}}e^{-i\frac{\w}{k}} \int_0^{\frac{k}{2}\bar{u}(u)}dse^{\frac{i\w
}{k}\left(s+1-\frac{k}{2}\bar{u}(u)\right)^2}
e^{-\frac{2\alpha s}{k}};& 0\leq u\leq u_0\\
&\\
\frac{1}{\sqrt{4\pi|\omega| }}\frac{-i\alpha}{\w+i\alpha}e^{-\alpha\bar{u}(u)}
 -\frac{1}{\sqrt{4\pi|\omega|
}}\frac{i\alpha}{\sqrt{A}\w+i\alpha}\left[ e^{-i\w V(u)}-e^{-i\w
V(u_0)}e^{-\alpha(\bar{u}(u)-\bar{u}(u_0))}
\right]& \\
& \\ -\frac{2\alpha}{k\sqrt{4\pi|\omega|
}}e^{-i\frac{\w}{k}}e^{-\alpha(\bar{u}(u)-\bar{u}(u_0))}
\int_0^{\frac{k}{2}\bar{u}(u_0)}dse^{\frac{i\w
}{4}\left(s+1-\frac{k}{2}\bar{u}(u_0)\right)^2} e^{-\frac{2\alpha
s}{k}};& u\geq u_0
\end{array}\right.
\end{eqnarray*}
and
\begin{eqnarray*}\hspace{-2mm}
\phi_{\w, L}^{trans}(u)=\left\{
\begin{array}{cc}
\frac{1}{\sqrt{4\pi|\omega| }}\frac{\w}{\w+i\alpha}e^{-i\w V(u)}
;&u\leq 0\\
&\\
\frac{1}{\sqrt{4\pi|\omega| }}e^{-i\w u}
 +\frac{1}{\sqrt{4\pi|\omega| }}\frac{-i\alpha}{\w+i\alpha}e^{-\alpha\bar{u}(u)}
 -\frac{2\alpha}{k\sqrt{4\pi|\omega|
}} \int^{\frac{k}{2}\bar{u}(u)}_0ds (s+1-\frac{k}{2}\bar{u}(u)
)^{2i\frac{\w}{k}} e^{-\frac{2\alpha s
}{k}};& 0\leq u\leq u_0\\
&\\\frac{1}{\sqrt{4\pi|\omega| }}\frac{-i\alpha}{\w+i\alpha} e^{-\alpha\bar{u}(u)}
+\frac{1}{\sqrt{4\pi|\omega|}}\frac{e^{-i\w u_0}}{\w+i\alpha\sqrt{A}}\left[ \w
e^{-i\frac{\w}{\sqrt{A}}(\bar{u}(u)-\bar{u}(u_0))}+i\alpha\sqrt{A}
e^{-\alpha(\bar{u}(u)-\bar{u}(u_0))}\right]& \\
& \\
-\frac{2\alpha}{k\sqrt{4\pi|\omega|
}}e^{-\alpha(\bar{u}(u)-\bar{u}(u_0))}
\int^{\frac{k}{2}\bar{u}(u_0)}_0ds (s+1-\frac{k}{2}\bar{u}(u_0)
)^{2i\frac{\w}{k}} e^{-\frac{2\alpha s }{k}} ;& u\geq u_0
\end{array}\right.
\end{eqnarray*}
Note that (as already advanced) in the case of perfect reflection,
that is when $\alpha\rightarrow \infty$, we get
\begin{eqnarray}
\phi_{\w, R}^{refl}(u)\rightarrow -\frac{1}{\sqrt{4\pi|\omega|
}}e^{-i\w V(u)}, \qquad \phi_{\w, L}^{trans}(u)\rightarrow 0,
\end{eqnarray}
and when the mirror is transparent, i.e., when $\alpha\rightarrow 0$, we have
\begin{eqnarray}
\phi_{\w, R}^{refl}(u)\rightarrow 0, \qquad \phi_{\w,
L}^{trans}(u)\rightarrow \frac{1}{\sqrt{4\pi|\omega| }}e^{-i\w u}.
\end{eqnarray}

We are interested in the particle production on the rhs of the
mirror, for this reason we must now obtain, for  $\w,\w'>0$
\begin{eqnarray}
 \beta^{R,R}_{\w,\w'}\equiv {({\phi_{\omega,R}^{out}}^*;\phi_{\omega',R}^{in})}^*,\quad \mbox{ and
 }\quad
\beta^{R,L}_{\w,\w'}\equiv
{({\phi_{\omega,R}^{out}}^*;\phi_{\omega',L}^{in})}^* .
\end{eqnarray}
In order to calculate these products we better choose the right null
infinity ${\mathcal J}^+_R$, because here the `out' modes acquire a
simple form, namely
\begin{eqnarray}
 \beta^{R,R}_{\w,\w'}= {({\phi_{\omega,R}^{out}}^*;\phi_{\omega',R}^{refl})}^*,\quad \mbox{ and
 }\quad
\beta^{R,L}_{\w,\w'}\equiv {({\phi_{\omega,R}^{out}}^*;\phi_{\omega',L}^{trans})}^*.
\end{eqnarray}
We start by calculating $\beta^{R,R}_{\w,\w'}=2i\int_{\R}du\,
\phi_{\omega,R}^{out}\,
\partial_u \phi_{\omega',R}^{refl}$, with the result
\begin{eqnarray}\label{eq2}
\beta^{R,R}_{\w,\w'}&\cong&
\frac{1}{2\pi\sqrt{\w\w'}}\frac{\alpha}{\w'+i\alpha}\left[1
-\frac{\alpha}{k} \int_A^1dx
x^{i{\w}/{k}-{1}/{2}}e^{-{2\alpha}(1-\sqrt{x})/k}\right]\nonumber\\&&\hspace{-1cm}
+\frac{\alpha}{2\pi k i\sqrt{\w\w'}}e^{-i{\w'}/{k}} \int_A^1dx
x^{i{\w}/{k}-{1}/{2}}e^{i{\w'x}/{k}}\left[1
-\frac{2\alpha}{k}
\int_0^{1-\sqrt{x}}e^{i{\w'}(s^2+2s\sqrt{x})/k}e^{-{2\alpha
s}/{k}}\right].
\end{eqnarray}
Now, provided that $\w'\ll \alpha$, Eq.~(\ref{eq2}) turns into
Eq.~(\ref{eq1}). As a consequence, we precisely obtain the same behavior
as for a perfectly reflecting mirror. However, in the case
$\alpha\ll \w'$,
\begin{eqnarray}
\beta^{R,R}_{\w,\w'}&\cong& \frac{\alpha}{2\pi k
i\sqrt{\w\w'}}e^{-i{\w'}/{k}}\left(i\frac{k}{\w'}\right)^{i{\w}/{k}+{1}/{2}}
\Gamma\left({1}/{2}+i{\w}/{k}\right),
\end{eqnarray}
and using the identity
$|\Gamma\left({1}/{2}+i{\w}/{k}\right)|^2=\frac{\pi}{\cosh\left(
 \pi{\w}/{k}\right)}$ (see \cite{as72}), we conclude that
\begin{eqnarray}\label{f}
\left|\beta^{R,R}_{\w,\w'}\right|^2&\cong& \frac{1}{2\pi
k\w}\left(\frac{\alpha}{\w'}\right)^2
\left(e^{2\pi{\w}/{k}}+1\right)^{-1}.
\end{eqnarray}

Finally, a simple but rather cumbersome calculation yields the
results
\begin{eqnarray}
\left|\beta_{\omega,\omega'}^{R,L}\right|^2\cong 0, \quad
\w'\ll\alpha,
\end{eqnarray}
and
\begin{eqnarray}
\left|\beta_{\omega,\omega'}^{R,L}\right|^2\sim \frac{1}{\w\w'}
{\mathcal O}\left[ \left(\frac{\alpha}{\w'}\right)^2\right], \quad
\alpha\ll\w'.
\end{eqnarray}
Note that in the case $\alpha\ll\w'$ we indeed obtain the nice
feature that  the number of created particles in the $\w$ mode,
together with the radiated energies, are both {\it finite}
quantities when $u_0\rightarrow \infty$, in perfect agreement with
the conclusions in \cite{n03}. More precisely, for a partially
transmitting mirror the number of produced particles in the $\w$
mode
$${\mathcal
N}_{\w}\equiv \int_0^{\infty}d\w'\left(\left|\beta_{\w,\w'}^{R,R}\right|^2 +
\left|\beta_{\w,\w'}^{R,L}\right|^2\right)
,$$
 is approximately
 $\int_0^{\infty}d\w'\left|\beta_{\w,\w'}^{R,R}\right|^2$.
In order to calculate this quantity, we split the domain
$[0,\infty)$ into two disjoints sets, $[0,k)$ and $[k,\infty)$,
respectively. In the second domain we can carry out the
approximation (\ref{f}) to obtain
\begin{eqnarray}\label{g}
\int_k^{\infty}d\w'\left|\beta_{\w,\w'}^{R,R}\right|^2\cong
\frac{1}{2\pi
\w}\left(\frac{\alpha}{k}\right)^2
\left(e^{2\pi\w/k}+1\right)^{-1}.
\end{eqnarray}
In the other domain, assuming that $k\ll 1$,   we have $\w'\ll 1$
and thus for incident waves of very low frequency the mirror behaves
like a perfect reflector. For this reason we can use the formula
(\ref{eq1}) and a simple calculation yields
\begin{eqnarray}
\int_0^{k}d\w'\left|\beta_{\w,\w'}^{R,R}\right|^2\sim {\mathcal
O}\left(\frac{ k^2}{\w(\w^2+k^2)}\right).
\end{eqnarray}
Thus, since $k\ll 1$, we conclude that the number of produced
particles in the $\w$ mode is approximately
\begin{eqnarray}
{\mathcal N}_\w\cong\frac{1}{2\pi \w}\left(\frac{\alpha}{k}\right)^2
\left(e^{2\pi\w/k}+1\right)^{-1}, \label{num41}
\end{eqnarray}
and the radiated energy ${\mathcal
E}\equiv\int_0^{\infty}d\w\hbar\w{\mathcal N}_{\w}$ is, also with
good approximation,
\begin{eqnarray}
{\mathcal E}\cong \frac{\hbar\alpha^2}{4\pi^2 k}\ln 2.
\end{eqnarray}
This completes the proof of all of the statements above.

It is appropriate to remark that there is a crucial difference with
the case $\w'\ll\alpha$, where the number of radiated particles in
the $\w$ mode diverges logarithmically with $u_0\rightarrow \infty$.
In this situation the physically relevant quantity is the number of
created particles in the $\w$ mode per unit time
$t_0\equiv \frac{1}{2}(u_0+V(u_0))\cong \frac{1}{2}u_0$. This dimensionless
quantity is finite and its value is given by \cite{n03,ha05}
\begin{eqnarray}\label{lim}
\lim_{t_0\rightarrow \infty}\frac{1}{t_0}{\mathcal N}_\w=\frac{1}{\pi}\left(
e^{2\pi\w/k}-1\right)^{-1}.
\end{eqnarray}

Finally, it is also interesting to calculate the {\it detector response
function} \cite{bd82}, namely ${\mathcal F}(\w)$, for an inertial DeWitt
detector following the trajectory $z=0$. This function is given by
\begin{eqnarray}
{\mathcal F}(\w)\equiv \int_{\R}dt\int_{\R}dt' e^{-i\w(t-t')}\langle 0,in|\phi(0,t) \phi(0,t')
|0,in\rangle,
\end{eqnarray}
where $|0,in\rangle$ denotes the `in' vacuum state.
This function is related to the average number of produced particles
in the $\w$-mode, through the relation
\cite{w85}
\begin{eqnarray}
{\mathcal F}(\w)=\frac{\pi}{\w}{\mathcal N}_\w.
\end{eqnarray}
Then, for a partially transmitting mirror, from Eq.~(\ref{num41}), we conclude that
\begin{eqnarray}
{\mathcal F}(\w)\cong\frac{1}{2\w^2}\left(\frac{\alpha}{k}\right)^2
\left(e^{2\pi\w/k}+1\right)^{-1}.
\end{eqnarray}
However, for a perfect reflecting mirror, this quantity diverges.
In this case the relevant function is the
{\it detector response function per unit time} \cite{t86}, namely ${\mathcal P}(\w)\equiv \lim_{t_0\rightarrow \infty}
\frac{1}{t_0}{\mathcal F}(\w)$. Using (\ref{lim}), one then obtains the Planckian spectrum
\begin{eqnarray}
{\mathcal P}(\w)=\frac{1}{\w}\left(
e^{2\pi\w/k}-1\right)^{-1}.
\end{eqnarray}

\section{Case of the Dirac field}

In this section we consider the Dirac equation, in $1+1$ dimensions,
 for a massless field
\begin{eqnarray}
\gamma^0\partial_t\psi+\gamma^1\partial_x\psi=0, \label{e44}
\end{eqnarray}
where here the Dirac matrices are (see \cite{cjjtw74})
\begin{eqnarray}
\gamma^0=\left(\begin{array}{cc} 0&1\\
1&0
\end{array}\right)\quad
\gamma^1=\left(\begin{array}{cc} 0&-1\\
1&0
\end{array}\right).
\end{eqnarray}
In the variables $(u,v)$, Eq.~(\ref{e44}) is
\begin{eqnarray}
\left(\begin{array}{cc} 0&1\\
0&0
\end{array}\right)\partial_u\psi+
\left(\begin{array}{cc} 0&0\\
1&0
\end{array}\right)\partial_v\psi=0,
\end{eqnarray}
and we conclude that its general solution can be written as
\begin{eqnarray}\psi(u,v)=\left(\begin{array}{c}F(u)\\G(v)
\end{array}\right).\end{eqnarray}

Consider now again the trajectory defined in Sect.~II. The
vector normal to the trajectory is
$n_{\mu}=\frac{1}{2\sqrt{V'(u)}}(V'(u),-1)$ and the current vector
is given by $j^{\mu}=2(|G|^2,|F|^2)$. Then, for a perfectly reflecting
mirror we must  impose that the normal component of the current vanishes
on the mirror (\cite{h78},\cite{r96}), that is, $j^{\mu}n_{\mu}=0$, and thus
 the  condition follows that
\begin{eqnarray}
V'(u)|G|^2-|F|^2=0.
\end{eqnarray}
From here we can calculate the corresponding family of `in´ and
`out' modes
\begin{eqnarray}
\psi_{\w,R}^{in}(u,v)\equiv\left[
\frac{1}{\sqrt{2\pi}}\left(\begin{array}{c}0\\1\end{array}\right)e^{-i\w
v}-\sqrt{\frac{V'(u)}{2\pi}}\left(\begin{array}{c}1\\0\end{array}\right)
e^{-i\w V(u)}\right]\theta(v-V(u)),
\end{eqnarray}
\begin{eqnarray}
\psi_{\w,R}^{out}(u,v)\equiv\left[
\frac{1}{\sqrt{2\pi}}\left(\begin{array}{c}1\\0\end{array}\right)e^{-i\w
u}-\sqrt{\frac{U'(v)}{2\pi}}\left(\begin{array}{c}1\\0\end{array}\right)
e^{-i\w U(v)}\right]\theta(v-V(u)).
\end{eqnarray}

In this case the beta Bogoliubov coefficient is given by
\begin{eqnarray}
\beta^{R,R}_{\w,\w'}\equiv (\psi_{\w,R}^{out
*};\psi_{\w',R}^{in})^*=\int(\psi_{\w,R}^{out })^t\psi_{\w',R}^{in},
\end{eqnarray}
where $(\psi_{\w,R}^{out })^t$ denotes the transposed of the vector
$\psi_{\w,R}^{out }$.

Performing the calculation in the right null future infinity domain, ${\mathcal
J}_R^+$, we obtain
\begin{eqnarray}
\beta^{R,R}_{\w,\w'}\cong \frac{1}{2\pi i \w'}-\frac{1}{2\pi k}
\int_0^{1-A}
ds(1-s)^{i{\w}/{k}-{1}/{2}}e^{-i{\w'}s/{k}}.
\end{eqnarray}
Then, as in \cite{ha05}, we easily get
\begin{eqnarray}
\beta^{R,R}_{\w,\w'}\cong \frac{1}{2\pi
k}e^{-i{\w'}/{k}}\left(\frac{ik}{\w'}\right)^{i{\w}/{k}+{1}/{2}}
\Gamma\left({1}/{2}+i{\w}/{k}\right),
\end{eqnarray}
and thus, since
$|\Gamma\left({1}/{2}+i{\w}/{k}\right)|^2=\frac{\pi}{\cosh\left(\pi{\w}/{k}
 \right)}$ (see \cite{as72}), we finally arrive to the anticipated
 result that
\begin{eqnarray}
\left|\beta^{R,R}_{\w,\w'}\right|^2\simeq \frac{1}{2\pi \w'
k}\left(e^{2\pi{\w}/{k}}+1\right)^{-1}.
\end{eqnarray}

\subsection{Partially reflecting, moving mirror}

We start with the two orthonormal basis
\begin{eqnarray}
\psi_{1,\w}(v)\equiv
\frac{1}{\sqrt{2\pi}}\left(\begin{array}{c}0\\1\end{array}\right)
\sqrt{\bar{v}'(v)}e^{-i\w \bar{v}(v)},\quad \psi_{2,\w}(u)\equiv
\frac{1}{\sqrt{2\pi}}\left(\begin{array}{c}1\\0\end{array}\right)
\sqrt{\bar{u}'(u)}e^{-i\w \bar{u}(u)},
\end{eqnarray}
where the functions $\bar{v}$ and $\bar{u}$ are defined in
Sect.~III. The normal component of the  mirror's current given by
each of these two functions are  $\frac{1}{2\pi}$ and
$-\frac{1}{2\pi}$, respectively.

It is interesting to observe the analogy  between the fermionic and
the scalar cases. Note that the current for a scalar field is
\begin{eqnarray}
j^{\mu}=i(\phi\partial_t\phi^*-\phi^*\partial_t\phi,
-\phi\partial_x\phi^*+\phi^*\partial_x\phi),
\end{eqnarray}
and if we choose the orthonormal basis:
\begin{eqnarray}
\phi_{1,\w}(v)\equiv \frac{1}{\sqrt{4\pi\w}} e^{-i\w
\bar{v}(v)},\quad \phi_{2,\w}(u)\equiv \frac{1}{\sqrt{4\pi\w}}
e^{-i\w \bar{u}(u)},
\end{eqnarray}
we obtain the same normal component of the current as for the respective
fermionic function. We can establish the following analogy
\begin{eqnarray}
\psi_{j,\w}\longleftrightarrow\phi_{j,\w},
\end{eqnarray}
and using this analogy we can construct the $\bar{g}^{in}$ modes in
the fermionic case as follows: simply replace in formulas (\ref{d})
and (\ref{e}) the functions $\phi_{j,\w}$ with the functions
$\psi_{j,\w}$, to obtain
\begin{eqnarray}\hspace{-1cm}
\bar{g}^{in}_{\w, R}(u,v)&=&
\frac{1}{\sqrt{2\pi}}s(\w)\left(\begin{array}{c}0\\1\end{array}\right)
\sqrt{\bar{v}'(v)}e^{-i\w \bar{v}(v)}\theta(u-U(v))\nonumber\\&&+
\frac{1}{\sqrt{2\pi}}\left[\left(\begin{array}{c}0\\1\end{array}\right)
\sqrt{\bar{v}'(v)}e^{-i\w \bar{v}(v)} +r(\w)
\left(\begin{array}{c}1\\0\end{array}\right)
\sqrt{\bar{u}'(u)}e^{-i\w \bar{u}(u)}\right]\theta(v-V(u))
\end{eqnarray}
and
\begin{eqnarray}\hspace{-1cm}
\bar{g}^{in}_{\w,
L}(u,v)&=&\frac{1}{\sqrt{2\pi}}\left[\left(\begin{array}{c}1\\0\end{array}\right)
\sqrt{\bar{u}'(u)}e^{-i\w \bar{u}(u)} +r(\w)
\left(\begin{array}{c}0\\1\end{array}\right)
\sqrt{\bar{v}'(v)}e^{-i\w
\bar{v}(v)}\right]\theta(u-U(v))\nonumber\\&&
+\frac{1}{\sqrt{2\pi}}s(\w)\left(\begin{array}{c}1\\0\end{array}\right)
\sqrt{\bar{u}'(u)}e^{-i\w \bar{u}(u)}\theta(u-U(v)).
\end{eqnarray}
As in the scalar case, the modes $\bar{g}^{in}$ are not the `in' modes
in the coordinates $(u,v)$. However using the fact that the modes
$\bar{g}^{in}$ constitute an orthonormal basis of the space of solutions, we
can obtain the following expression of the `in' modes:
\begin{eqnarray}
\psi^{in}_{\w,k}=\int_{\R}d\w'(\bar{g}^{in}_{\w',k};\psi^{in}_{\w,k}) \bar{g}^{in}_{\w',k}.
\end{eqnarray}
If we choose the coefficients
 $r(w)=\frac{-i\alpha}{\w+i\alpha}$ and $s(w)=\frac{\w}{\w+i\alpha}$,
with $\alpha\geq 0$, on the right side of mirror we have
\begin{eqnarray}
\psi^{in}_{\w,R}(u,v)=\frac{1}{\sqrt{2\pi
}}\left(\begin{array}{c}0\\1\end{array}\right) e^{-i\omega v}+
\psi_{\w, R}^{refl}(u), \qquad \psi^{in}_{\w,L}(u,v)=\psi_{\w,
L}^{trans}(u),
\end{eqnarray}
where
\begin{eqnarray*}\hspace{-5mm}
\psi_{\w,
R}^{refl}(u)=\frac{1}{\sqrt{2\pi}}\left(\begin{array}{c}1\\0\end{array} \right)\left\{
\begin{array}{cc}
\frac{-i\alpha\sqrt{V'(u)}}{\w+i\alpha}
e^{-i\w V(u)};&u\leq 0\\
&\\
 \frac{-i\alpha\sqrt{\bar{u}'(u)}}{\w+i\alpha}
 e^{-\alpha\bar{u}(u)}&\\
 &\\
 -\frac{2\alpha e^{-i\frac{\w}{k}}\sqrt{\bar{u}'(u)} }{k}
 \int_0^{\frac{k}{2}\bar{u}(u)}ds\sqrt{s+1-\frac{k}{2}\bar{u}(u)}
 e^{\frac{i\w}{k}\left(s+1-\frac{k}{2}\bar{u}(u)\right)^2}
e^{-\frac{2\alpha s}{k}};& 0\leq u\leq u_0\\
&\\
\frac{-i\alpha\sqrt{\bar{u}'(u)}}{\w+i\alpha} e^{-\alpha\bar{u}(u)}
 -
 \frac{i\alpha\sqrt{V'(u)}}{\sqrt{A}\w+i\alpha}\left[ e^{-i\w V(u)}-e^{-i\w
V(u_0)}e^{-\alpha(\bar{u}(u)-\bar{u}(u_0))}
\right]& \\
& \\-\frac{2\alpha e^{-i\frac{\w}{k}} \sqrt{\bar{u}'(u)}e^{-\alpha(\bar{u}(u)-\bar{u}(u_0))} }{k}
 \int_0^{\frac{k}{2}\bar{u}(u_0)}ds\sqrt{s+1-\frac{k}{2}\bar{u}(u_0)}
 e^{\frac{i\w}{k}\left(s+1-\frac{k}{2}\bar{u}(u_0)\right)^2}
e^{-\frac{2\alpha s}{k}} ;& u\geq u_0
\end{array}\right.
\end{eqnarray*}
and
\begin{eqnarray*}\hspace{-5mm}
\psi_{\w,
L}^{trans}(u)=\frac{1}{\sqrt{2\pi}}\left(\begin{array}{c}1\\0\end{array} \right)\left\{
\begin{array}{cc}
\frac{\w\sqrt{V'(u)}}{\w+i\alpha}
e^{-i\w V(u)};&u\leq 0\\
&\\
 \frac{-i\alpha\sqrt{\bar{u}'(u)}}{\w+i\alpha}
 e^{-\alpha\bar{u}(u)}+e^{-i\w u}&\\
 &\\
 -\frac{2\alpha \sqrt{\bar{u}'(u)} }{k}
 \int_0^{\frac{k}{2}\bar{u}(u)}ds(s+1-\frac{k}{2}\bar{u}(u))^{2i\w/k-1/2}
e^{-2\alpha s/k};& 0\leq u\leq u_0\\
&\\
\frac{-i\alpha\sqrt{\bar{u}'(u)}}{\w+i\alpha} e^{-\alpha\bar{u}(u)}
 +
 \frac{e^{-i\w u_0}}{\w+i\alpha\sqrt{A}}\left[\w e^{-i\frac{\w}{\sqrt{A}}(\bar{u}(u)-\bar{u}(u_0))}
+i\alpha\sqrt{A}
e^{-\alpha\sqrt{A}(\bar{u}(u)-\bar{u}(u_0))}
\right]&
\\
& \\-\frac{2\alpha  \sqrt{\bar{u}'(u)}}{k}
 e^{-\alpha(\bar{u}(u)-\bar{u}(u_0))} \int_0^{\frac{k}{2}\bar{u}(u_0)}
ds(s+1-\frac{k}{2}\bar{u}(u_0))^{2i\w/k-1/2}
e^{-2\alpha s/k}
.& u\geq u_0
\end{array}\right.
\end{eqnarray*}

Note that in the case of perfect reflection, that is, when
$\alpha\rightarrow \infty$, we have
\begin{eqnarray}
\psi_{\w, R}^{refl}(u)\rightarrow
-\sqrt{\frac{V'(u)}{2\pi}}\left(\begin{array}{c}1\\0\end{array}\right)e^{-i\w
V(u)}, \qquad \psi_{\w, L}^{trans}(u)\rightarrow 0,
\end{eqnarray}
and when the mirror is transparent, i.e., when $\alpha\rightarrow 0$,
it turns out that
\begin{eqnarray}
\psi_{\w, R}^{refl}(u)\rightarrow 0, \qquad \psi_{\w,
L}^{trans}(u)\rightarrow \frac{1}{\sqrt{2\pi
}}\left(\begin{array}{c}1\\0\end{array}\right)e^{-i\w u}.
\end{eqnarray}
To calculate the production of particles on the rhs of the mirror we
must obtain $\beta^{R,R}_{\w,\w'}\equiv (\psi_{\w,R}^{out
*};\psi_{\w',R}^{in})^*$ and $\beta^{R,L}_{\w,\w'}\equiv (\psi_{\w,R}^{out
*};\psi_{\w',L}^{in})^*$. We start by calculating $\beta^{R,R}_{\w,\w'}$.
Choosing the right null infinity region ${\mathcal J}_R^+$, we have
$\beta^{R,R}_{\w,\w'}=\int_{R}(\psi_{\w,R}^{out})^t\psi_{\w',R}^{refl}du$,
then, for $1\ll{\w'}/{k}\ll e^{ku_0}$ and $1\ll{\w'}/{\w}\ll
e^{ku_0}$, we get
\begin{eqnarray}
\beta^{R,R}_{\w,\w'}&\cong&
\frac{1}{2\pi}\frac{\alpha}{\w'+i\alpha}\frac{1}{\w+\w'}\nonumber\\&&
-\frac{\alpha e^{-i\w'/k}}{k^2\pi}\int_{A}^1dx
x^{i\w/k-3/4}\int_0^{(1-\sqrt{x})}ds\sqrt{s+\sqrt{x}} e^{\frac{i\w'}{k}(s+\sqrt{x})^2}e^{-2\alpha
s/k}.
\end{eqnarray}
Integrating by parts with respect to the $s$ variable, we  obtain
\begin{eqnarray}
\beta^{R,R}_{\w,\w'}\cong -\frac{\alpha}{2\pi i \w k}\left(\frac{i
k}{\w'}\right)^{i\w/k+1}\Gamma\left(1+i\w/k\right),
\end{eqnarray}
and thus, in the case of a partially reflecting mirror, it turns out
that
\begin{eqnarray}
\left|\beta^{R,R}_{\w,\w'}\right|^2\cong\frac{1}{2\pi \w
k}\left(\frac{\alpha}{\w'}\right)^2\left(e^{2\pi\w/k}-1\right)^{-1}.
\end{eqnarray}
Finally, a simple but rather cumbersome calculation yields the
result we were looking for
\begin{eqnarray}
\left|\beta_{\omega,\omega'}^{R,L}\right|^2\sim \frac{1}{\w\w'}
{\mathcal O}\left[ \left(\frac{\alpha}{\w'}\right)^2\right].
\end{eqnarray}

\section{Conclusions}
In this paper we have studied in detail the creation of scalar
massless particles in a two-dimensional Minkowski space-time (the
Davies-Fulling theory) and specifically for the case of a
semi-transparent mirror, which reflects low frequency modes but is
transparent to high enough frequencies, being the reflection and
transmission coefficients analytic functions of the frequency. The
considered mirror is initially at rest, then accelerates, during
some finite time, along a trajectory that simulates a black hole
collapse (as defined by Walker \cite{w85}, and Carlitz and Willey
\cite{cw87}), and finally rests moving with constant velocity.

When the reflection and transmission coefficients are those in the
model proposed by Barton, Calogeracos, and Nicolaevici
\cite{bc95,n01,c02a}, namely $r(w)=-i\alpha/(\w+i\alpha)$ and
$s(w)=\w/(\w+i\alpha)$, with $\alpha\geq 0$, the Bogoliubov
coefficients on the back side of the mirror could be computed exactly.
This has allowed us to rigorously prove that, when $\alpha$ is very
large (the case of an ideal, perfectly reflecting mirror) a thermal
emission of scalar massless particles obeying Bose-Einstein
statistics is radiated from the mirror (a black body radiation), in
accordance with previous results in the literature. Moreover, we
have also seen that when $\alpha$ is finite (the case of a
semi-transparent mirror, that is, a physically realistic situation)
the surprising result is obtained that the thermal emission of
scalar massless particles obeys  Fermi-Dirac statistics. We have
also shown in detail that the reverse change of statistics takes
place in a bidimensional fermionic model for massless particles,
namely, that the Fermi-Dirac statistics for the completely
reflecting situation gives rise to the Bose-Einstein statistics for
the case of a semi-transparent, physical mirror.

The results we have obtained are absolutely solid---they do not hang
on a perturbative expansion or approximation of any sort. The
physical reason for this surprising change of statistics may be
found in the fact that the form of the spectrum is actually
determined {\it not} through the statistics of the field but rather
by the specific trajectory of the mirror and by its interaction with
the radiation field. The same kind of phenomenon occurs in the case
of an electric charge following the trajectory
$v=\frac{1}{k}(1-e^{-ku})$. When the radiation field has spin $1$,
the radiation emitted by the charge obeys Bose-Einstein statistics,
but when a scalar charge, and consequently an scalar radiation
field, is considered, the emitted radiation will obey  Fermi-Dirac
statistics \cite{nr95}.

Another situation where this kind of features occurs, is when one
measures the spectrum of an scalar field using a DeWitt detector
\cite{dw75,bd82} which follows a uniformly accelerated world-line
in Minkowski space-time. In this case when the dimension of the
space-time is even the Bose-Einstein statistics is obtained.
However, when the dimension is odd, precisely the reverse change of
statistics occurs in the emitted radiation \cite{t86,oo86,u86,te99}.

\vspace{1cm}

\noindent {\bf Acknowledgements.} This investigation is partly based
on work done while on leave at the Department of Physics and
Astronomy, Dartmouth College, 6127 Wilder Laboratory, Hanover, NH
03755, USA. This research was supported by MEC (Spain), projects
MTM2005-07660-C02-01 and FIS2006-02842, and by AGAUR
(Gene\-ra\-litat de Ca\-ta\-lu\-nya), grant 2007BE-1003 and contract
2005SGR-00790.

\vspace{1cm}

\section{Appendix}
We will here derive in detail the total number of produced particles
and their energy. First, we start with the case of a perfectly
reflecting mirror. Assuming that the mirror's velocity converges
fast enough to some constant, when $|u|\rightarrow\infty$, we have
\begin{eqnarray}
\beta_{\w,\w'}^{R,R}=\frac{1}{2\pi}\sqrt{\frac{\w}{\w'}}
\int_{\R}due^{-i\w u}e^{-i\w' V(u)}
\end{eqnarray}
and, integrating by parts, we get
\begin{eqnarray}
\beta_{\w,\w'}^{R,R}=-\frac{1}{2\pi i}\sqrt{\w\w'}
\int_{\R}du\frac{V''(u)}{(\w+\w' V'(u))^2} e^{-i\w u}e^{-i\w' V(u)}.
\end{eqnarray}
For simplicity,  we will assume that the mirror's acceleration is
discontinuous at the point $u=a$. After another integration by
parts, we obtain
\begin{eqnarray}
\beta_{\w,\w'}^{R,R}&&=-\frac{1}{2\pi }\sqrt{\w\w'} \frac{1}{(\w+\w'
V'(a))^3} e^{-i\w a}e^{-i\w' V(a)}(V''(a^-)-V''(a^+)) \nonumber\\&&
+\frac{1}{2\pi }\sqrt{\w\w'} \int_{\R}du\left[\frac{V'''(u)}{(\w+\w'
V'(u))^3}- \frac{3\w' (V''(u))^2}{(\w+\w' V'(u))^4}\right] e^{-i\w
u}e^{-i\w' V(u)}.
\end{eqnarray}
From this expression, if we further assume that the mirror's
trajectory is asymptotically inertial,  that is $V'(u)>0, \
\forall u\in \R$ (see for example \cite{w85}), it also follows that
$\left|\beta_{\w,\w'}^{R,R}\right|^2$ and $\hbar
\w\left|\beta_{\w,\w'}^{R,R}\right|^2$ are integrable functions in
the domain $[0,\infty)^2\setminus [0,1]^2$.

Now, we are interested in the production of particles in the
infrared domain, that is, we want to calculate
$\left|\beta_{\w,\w'}^{R,R}\right|^2$ in $[0,1]^2$.
% We also assume
%that the trajectory considered above is static when $u\geq u_0$ (In
%the following discussion we only need that the initial and final
%mirror's velocity are the same).
%To obtain the value of the Bogolubov coefficient we decompose the
%integral in three parts $A\equiv\int_{-\infty}^0...$,
%$B\equiv\int_0^{u_0}...$ and $C\equiv\int_{u_0}^{\infty}...$. The
%first and third integrals can be calculated explicitly, the second
%can be written as follow
We write the Bogoliubov coefficient as follows
\begin{eqnarray}
\beta_{\w,\w'}^{R,R}=\frac{1}{2\pi}\sqrt{\frac{\w}{\w'}}
\int_{\R}due^{-i(\w+B\w') u}e^{-i\w' (V(u)-Bu)},
\end{eqnarray}
with $B>0$. After integration by parts, we obtain
\begin{eqnarray}
\beta_{\w,\w'}^{R,R}=-\frac{1}{2\pi}\frac{\sqrt{\w\w'}}{\w+B\w'}
\int_{\R}du(V'(u)-B)e^{-i\w u}e^{-i\w' V(u)},
\end{eqnarray}
and thus, if the function $|V'(u)-B|$ is integrable in $\R$ for some
$B>0$, it can be deduced that  $\left|\beta_{\w,\w'}^{R,R}\right|^2$
and $\hbar \w\left|\beta_{\w,\w'}^{R,R}\right|^2$ are integrable
functions in the domain $[0,1]^2$.

An example of this kind of trajectories is
\begin{eqnarray}
V(u)=\left\{\begin{array}{cc} Bu & u\leq 0 \\
V(u) & 0\leq u\leq u_0\\
V(u_0)+B(u-u_0) & u\geq u_0. \end{array}\right.
\end{eqnarray}
However, if we are only interested in the convergence of the function
$\hbar \w\left|\beta_{\w,\w'}^{R,R}\right|^2$ in the domain
$[0,1]^2$, we only need trajectories that satisfy
\begin{eqnarray}\label{c}
\int_{-\infty}^0 du|V'(u)-B_1|<\infty \ \ \ \mbox{ and }\ \
\int^{\infty}_0du|V'(u)-B_2|<\infty
\end{eqnarray}
 for some non-negative constants $B_1$ and
$B_2$ (here it is important to remark that one of  these constants
can be zero, that is, it is not worthwhile that the trajectory be
asymptotically inertial). To prove that statement, we write
\begin{eqnarray}
\beta_{\w,\w'}^{R,R}=\frac{1}{2\pi}\sqrt{\frac{\w}{\w'}}\left[
\int_{-\infty}^0due^{-i(\w+B_1\w') u}e^{-i\w'
(V(u)-B_1u)}+\int_0^{\infty}due^{-i(\w+B_2\w') u}e^{-i\w'
(V(u)-B_2u)}\right],
\end{eqnarray}
and assume, for simplicity, that $V(0)=0$. After integration by
parts, we get the expression
\begin{eqnarray}
\beta_{\w,\w'}^{R,R}&=&-\frac{1}{2\pi}\sqrt{\w\w'}
\frac{B_1-B_2}{(\w+B_1\w')(\w+B_2\w')}-\frac{1}{2\pi} \frac{\sqrt{\w\w'}}{\w+B_1\w'}
\int_{-\infty}^0du(V'(u)-B_1)e^{-i\w u}e^{-i\w'
V(u)}\nonumber\\&&-\frac{1}{2\pi}\frac{\sqrt{\w\w'}}{\w+B_2\w'}
\int_{0}^{\infty}du(V'(u)-B_2)e^{-i\w u}e^{-i\w' V(u)},
\end{eqnarray}
which already proves the assertion.

In conclusion, we have here demonstrated that, for asymptotically
inertial trajectories with continuous velocity, the radiated energy
is indeed finite. However it is also possible that an infinite
production of particles with very low frequency could take place (an
infrared divergence). To remove this divergence we must just assume that
the initial and the final mirror velocities are the same.

For completeness, we should comment on the very interesting process
of particle creation, for the case of a partially
transmitting mirrors. In this situation, at high enough frequency the
mirror behaves as transparent, and then there is no particle
production, with independence of the mirror's trajectory. On the
other hand, at very low frequencies the mirror behaves as a perfect
reflector, and then we have the same kind of infrared problems as
for the perfectly reflecting case. As a consequence, if we are only
interested in the case when the radiated energy is finite, we must
restrict ourselves to consider trajectories with a continuous
velocity $V'(u),\ \forall u\in\R$, which fulfill the condition
(\ref{c}) as, for instance, the non-asymptotically inertial trajectory:
\begin{eqnarray}
v=V(u)\equiv\left\{\begin{array}{ccc}
u&\mbox{if}& u\leq 0\\
&&\\
\frac{1}{k}(1-e^{-ku})&\mbox{if}& u\geq 0.
\end{array}\right.\end{eqnarray}
This case can be discussed along the same lines above.


\begin{thebibliography}{99}

\bibitem{fd76} S.A. Fulling and P.C.W. Davies, Proc. Roy. Soc. Lond.
{\bf  A348}, 393 (1976).

\bibitem{fd77} P.C.W. Davies and S.A. Fulling, Proc. Roy. Soc. Lond.
{\bf  A356}, 237 (1977).

\bibitem{he06} J. Haro and E. Elizalde, Phys. Rev. Lett. {\bf 97}, 130401 (2006).

\bibitem{he06a} J. Haro and E. Elizalde, Phys. Rev. {\bf D76}, 065001 (2007).

\bibitem{w85}  W.R. Walker, Phys. Rev.  {\bf  D31}, 767 (1985).

\bibitem{cw87}  R.D. Carlitz and R.S. Willey, Phys. Rev.
{\bf  D36}, 2327 (1987).

\bibitem{c02b} A. Calogeracos, J. Phys. A: Math. Gen.
{\bf  35}, 3435 (2002).

\bibitem{n03}  N. Nicolaevici, J. Phys. A: Math. Gen.
{\bf  36}, 7667 (2003).

\bibitem{ha05} J. Haro, J. Phys. A: Math. Gen.
{\bf 38}, L307 (2005).

\bibitem{bc95} G. Barton and A. Calogeracos, Ann. Phys. (NY)
{\bf  238}, 227 (1995).

\bibitem{n01}  N. Nicolaevici, Class. Quantum Grav.
{\bf  18}, 619 (2001); {\bf  18}, 2895 (2001).

\bibitem{c02a}  A. Calogeracos, J. Phys. A: Math. Gen.
{\bf  35}, 3415 (2002).

\bibitem{12p} J. Haro and E. Elizalde, J. Phys. {\bf A41}, 032002 (2008).

\bibitem{nr95} A.I. Nikishov and V.I. Ritus, J.E.T.P.
{\bf  81}, 615 (1995).

\bibitem{t86} S. Takagi,  Prog. Theo. Phys. Supp. {\bf 88}, 1 (1986).

\bibitem{bd82} N.D. Birrell and C.P.W. Davies, {\it  Quantum Fields
in Curved Space} (Cambridge University Press, Cambridge, 1982).

\bibitem{dw75} B.S. DeWitt, Phys. Reports {\bf 19}, 295 (1975).

\bibitem{as72} M. Abramowitz and I.A. Stegun,
{\it Handbook of Mathematical Functions} (Dover, New York, 1972).

\bibitem{op01} N. Obadia  and R. Parentani, Phys. Rev. {\bf  D64}, 044019 (2001).

\bibitem{jr91} M.-T. Jaekel and S. Reynaud,  J. Phys. {\bf  I1}, 1395 (1991).

\bibitem{cjjtw74} A. Chodos, L.R. Jaffe, K. Johnson, C.B. Thorn and V.F. Weisskopf, Phys. Rev.
{\bf  D9}, 3471 (1974).

\bibitem{h78}  M. Horibe, Prog. Theor. Phys. {\bf  61}, 661 (1979).

\bibitem{r96}  V.I. Ritus, J. Exp. Theor. Phys. {\bf  83}, 282 (1996).

\bibitem{oo86} H. Ooguri, Phys. Rev. {\bf  D33}, 3573 (1986).

\bibitem{u86} W.G. Unruh, Phys. Rev. {\bf  D34}, 1222 (1986).

\bibitem{te99} H. Terashima, Phys. Rev. {\bf  D60}, 084001 (1999).

\end{thebibliography}
\end{document}